\documentclass[twocolumn,prb,amsmath,amssymb,amsfonts,superscriptaddress,floatfix,aps,showpacs,notitlepage]{revtex4-2}
\usepackage{graphicx}
\usepackage{float}
\usepackage{rotating}
\usepackage[utf8]{inputenc} 
\usepackage{longtable}
\usepackage{dcolumn}
\usepackage{bm}
\usepackage{color}
\usepackage{xcolor}

\usepackage[caption=false]{subfig}

\usepackage{dcolumn}
\usepackage{bm}
\usepackage[normalem]{ulem}
\usepackage{multirow}

\newcommand{\wideunderline}[2][2em]{%
	\underline{\makebox[\ifdim\width>#1\width\else#1\fi]{#2}}%
}

\newcommand{\be}{\begin{equation}}
\newcommand{\ee}{\end{equation}}
\newcommand{\bea}{\begin{eqnarray}}
\newcommand{\eea}{\end{eqnarray}}

\DeclareUnicodeCharacter{2212}{-}

\begin{document}
	\title{Engineering the electronic and magnetic properties of monolayer TiS$_2$ through systematic transition-metal doping}
	
\affiliation{Department of Physics, University of Guilan, 41335-1914, Rasht, Iran}	

\author{Sara Asadi Toularoud}

\author{Hanif Hadipour}
\email{hanifhadipour@gmail.com}

\author{Hamid Rahimpour Soleimani}

	\date{\today}

	
	\begin{abstract}

Layered materials that exhibit magnetic ordering in their pristine form are very rare. Several standard approaches, such as adsorption of atoms, introduction of point defects, and edge engineering, have been developed to induce magnetism in two-dimensional materials.
In this way, we investigate the electronic and magnetic properties of monolayer TiS$_2$ doped with 3$d$ transition metals (TMs) atoms in both octahedral 1T and trigonal prismatic 1H structures using first-principles calculations. In its pristine form, TiS$_2$ is a non-magnetic semiconductor. The bands near the Fermi energy primarily exhibit $d$ orbital characters, and due to the presence of ideal octahedral and trigonal arrangements, they are well separated from other bands with $p$ character.
Upon substituting 3$d$-TM atoms in both structures, a variety of electronic and magnetic phases emerge, including magnetic semiconductor, magnetic half-metal, non-magnetic metal, and magnetic metal. Chromium exhibits the largest magnetic moment in both the 1T and 1H structures. The 1T structure shows a slightly higher magnetic moment of 3.419 $\mu_B$ compared to the 1H structure 3.138 $\mu_B$, attributed to the distorted octahedral structure of the 1T structure. Unlike pristine TiS$_2$, the deficiency in saturation of neighboring S atoms in the presence of impurities leads to the proximity of energy levels of $d$ and $p$ states, resulting in unexpectedly sizable magnetic moments.
Another interesting case is Cobalt, which leads to a magnetic moment of approximately 0.805 $\mu_B$ in the 1H structure, while the Co exhibits a non-magnetic state in the 1H structure. These materials demonstrate a high degree of tunability and can be optimized for various magnetic applications.

	\end{abstract}
	
\pacs{73.22.-f, 68.65.−k, 71.35.−y, 71.10.−w}
	
\maketitle

	\section{Introduction}\label{sec1}

The synthesis of graphene has been highly successful, sparking significant interest in other two-dimensional materials because of their unique properties and wide range of potential applications \cite{Glavin,Huang,Tan}. This interest has led to the discovery of numerous two-dimensional materials with exceptional properties and diverse applications. Among these structures, transition metal dichalcogenides (TMDs) are particularly intriguing. TMDs have the chemical formula MX$_2$, where M represents a transition metal (TM) and X denotes a chalcogen (S, Se, Te, or combinations) \cite{Arul,Liu,Joseph,Ramezani}. The bonds between layers in TMDs are van der Waals, allowing for the production of single or multiple layers through various methods. For example, techniques like chemical vapor deposition (CVD) \cite{Zhan2012,Keng} and mechanical exfoliation \cite{Kyungnam,Islam} are commonly employed. One notably fascinating TMD is titanium disulfide TiS$_2$. In its bulk form, TiS$_2$ behaves as a semimetal with pronounced anisotropy. However, when exfoliated to a few layers, TiS$_2$ displays a range of intriguing phenomena. TiS$_2$ has attracted significant attention due to its potential applications in hydrogen storage \cite{Ren}, Lithium-Ion batteries \cite{Paul,Wei,Pham}, favorable thermal properties in thermoelectric materials \cite{Zavjalov,Xiaodong,Salleh}, and its useful optical characteristics \cite{Ziliang,Dorothy,Xiao}. Additionally, the examination of magnetic properties in TMDs is often conducted alongside investigations into these compelling physical properties.

Layered materials that naturally display magnetic ordering in their pristine form are rare. Among the vast number of 2D materials, only CrX$_3$(X=Br,I) \cite{Zhang-2019,Yekta,Siena,Seyler}, Cr$_2$Ge$_2$Te$_6$ \cite{Gong}, M$_2$C (M=Cr, Fe) \cite{Yasin2019}, and VX$_2$ (X=Se,Te) \cite{Duvjir, Bonilla, Karbala} have been experimentally observed to exhibit magnetic orderings.
Therefore, various innovative techniques have been employed to enhance or induce magnetic properties in 2D systems such as adsorption of atoms \cite{Sasioglu-1}, introduction of point defects \cite{Hadipour,Esquinazi,Maghool}, and edge engineering\cite{Bagherpour}.  In the case of TMDs, Yan Zhu et al. \cite{Yan2022} recently published a study on the influence of vacancy defects on the magnetic characteristics of VSe$_2$. Through a systematic investigation, the magnetic properties of PtSe$_2$ were analyzed by removing platinum and selenium atoms. The findings revealed that removing the platinum atom resulted in magnetization of the system, while the elimination of the selenium atom had no impact on the system's magnetism \cite{Avsar}. Moreover, researchers explored the effects of vacancies in Mo atoms on the magnetization of MoSe$_2$, MoTe$_2$, and WS$_2$. The research revealed that removing Mo atoms solely in MoSe$_2$ led to the generation of magnetic moments \cite{Ma}.

\begin{figure*}[t]
\begin{center}
	\includegraphics[width=150mm]{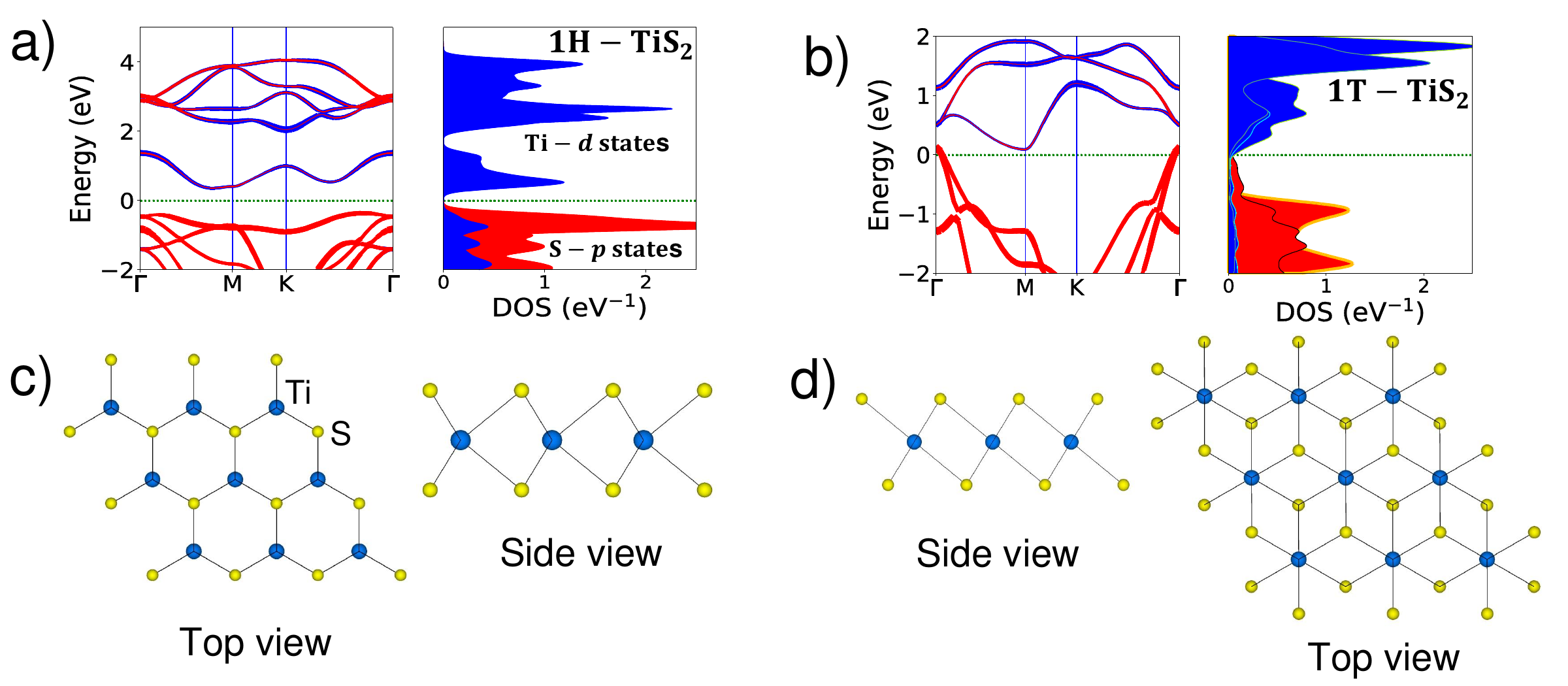}
\end{center}
\vspace*{-0.5cm} \caption{Orbital projected band structure and density of state (PDOS) for (a) 1H structure of single layer of TiS$_2$ and (b) 1T structure of single layer of TiS$_2$. Top and side view crystal structure of (c) 1H structure and (d) 1T structure. The Fermi level is set to zero energy.}
\label{fig:subm1}
\end{figure*}

Another approach to induce the magnetic properties of materials involves incorporating other TMs, particularly $3d$ elements, as dopants in the structure of TMDs.  Replacing the Mo/W atom with Mn, Fe, Co, Ni, Cu, or Zn atoms stabilizes the Mo(W)S$_2$ monolayer and introduces magnetic properties to the compound. However, elements like Sc, Ti, V, and Cr, which have fewer valence electrons than Mo, do not induce any magnetic moment \cite{Shanshan,Yungang-2013}. Introducing certain $3d$ atoms into the ZrS$_2$ monolayer instead of the Zr atom disrupts the symmetry of the upper and lower spins, resulting in a blend of magnetic properties \cite{Baishun,Maghool}. In another theoretical study by Yang et al., replacing W atoms with $3d$ elements like Zn, Cu, Ni, Co, Fe, and Mn in the single-layer of WS$_2$ triggers the emergence of magnetism in the system. Notably, the Ni atom generates the most significant amount of magnetism \cite{Yi}. Studies on the substitution of $3d$ TMs and the investigation of magnetism and its origins in various TMDs, including CrS$_2$, HfS$_2$, and WSe$_2$, have been documented \cite{Jianmin2015,Wang2016,Dinh}. Yue et al. explored magnetic anisotropies in Mn, Fe, and Co-doped monolayer MoS$_2$ \cite{Yunliang2020}.  Furthermore, Chowdhury et al. \cite{Sayantika} conducted a comprehensive investigation into the electronic structure of $3d$, $4d$, and $5d$ TM-doped WSe$_2$ monolayers. Several studies have analyzed the magnetism of materials in relation to their respective structural phases. For example, Ataca et al. conducted a systematic examination of the magnetism of VX$_2$ (X=O, S, Se, Te) in two structures, octahedral 1T and trigonal prismatic 1H, revealing that this material maintains its magnetic properties across both structures \cite{Can}.

While research on the magnetic properties of TMDs has been conducted, studies on the magnetization of TiS$_2$ monolayers are limited. Titanium disulfide is the lightest compound among the Group-IV TMDs. In a single-layer, TiS$_2$ has a band gap of 0.05 eV (0.9 eV) in 1T (1H) structure [see Figs.~\ref{fig:subm1}(a) and   ~\ref{fig:subm1}(b)]. The spatial and structural arrangement of these two structures is illustrated in Figs.~\ref{fig:subm1}(c) and   ~\ref{fig:subm1}(d). In both structures, a layer of titanium atoms is sandwiched between two layers of sulfur atoms. In 1T-TiS$_2$, each titanium atom is surrounded by six sulfur atoms in an octahedral arrangement, while in 1H-TiS$_2$, the coordination is trigonal prismatic.  

Studying the substitution of 3$d$ TM in non-magnetic semiconductor TiS$_2$ provides a fundamental insight into the origin of magnetic properties in two-dimensional systems. The variation in crystal field splitting of $d$ levels induced by neighboring chalcogen S atoms in the 1H and 1T lattices results in distinct correlated subspaces. This, in turn, elucidates the impact of orbital orientation on diverse magnetic ordering phenomena. This research delves into the electronic and magnetic characteristics of TiS$_2$ with the inclusion of  V, Cr, Mn, Fe, Co, Ni, Cu, and Zn atoms as impurities in both the 1H and 1T phases using first-principles density functional theory (DFT).
Our calculations suggest that Cr impurity is the most suitable candidates for magnetizing the system, exhibiting magnetic moments of around 3.1-3.5 $\mu_B$. The local magnetization of early TM impurities such as V, Cr, Mn are higher in the 1T structure than in the 1H one. In the case of Cr/Mn-doped TiS$_2$ in the both structures, there are  $d_{z^2}$ and $d_{x^2−y^2}$/$d_{xy}$ impurity levels  with a narrow bandwidth of approximately 0.4 eV. Due to the large DOS of Mn and Cr atoms
at the Fermi energy ($E_F$) in both 1T and 1H structures, they can be unstable to the magnetic ordering, which is
reasonably consistent with our results of spin-polarized total energy calculations and the sizable
magnetic moments.
The average magnetization of the six S atoms positioned as nearest neighbors to the impurity, as well as the magnetization of the closest Ti atom to the impurity, ranges between -1.3 to 0.7 $\mu_B$. This suggests a delocalized magnetization that spreads across multiple atoms. In the 1H structure, the neighboring atoms exhibit significantly higher magnetization compared to the 1T structure, indicating a smaller charge transfer from the impurities to their neighbors in the 1H phase.

	\section{Computational method}\label{sec2}

For DFT calculation we have used pseudopotential Quantum Espresso code \cite{espresso}
based on plane wave basis set within the generalized gradient approximation in
the Perdew-Burke-Ernzerhof (PBE) parameterization \cite{Perdew}. Simulation of both pristine and impurity doped TiS$_2$ unit cells is based on the slab model having a 25 $\AA$ vacuum
separating slabs. 
First, the structural properties after the ionic relaxations such as lattice
parameters, $x$, $y$, and $z$ component of the Ti, TM, and S atoms in crystal coordinates
are extracted. The uniform $k$-point grids of 36 $\times$ 36 $\times$ 1 (12 $\times$ 12 $\times$ 1) are used for the selfconsistent field calculations of pristine TiS$_2$ (TM-doped TiS$_2$). The Kinetic energy cut-offs for the
wavefunctions and the charge density are 850 and 8500 eV, respectively. For
each systems, the Broyden-Fletcher-Goldfarb-Shanno quasi-Newton algorithm
is used to relax the internal coordinates of atoms and possible
distortions with convergence threshold on forces for ionic minimization as small
as 10$^{-4}$ eV$\AA^{-1}$. 
We construct a 4 $\times$ 4 $\times$ 1
supercell based on the primitive cell of Figs.~\ref{fig:subm1}(c) and  ~\ref{fig:subm1}(d) to investigate the 3$d$-TM substitution in TiS$_2$.

\begin{figure}[t]
	\centering
	\includegraphics[width=84mm]{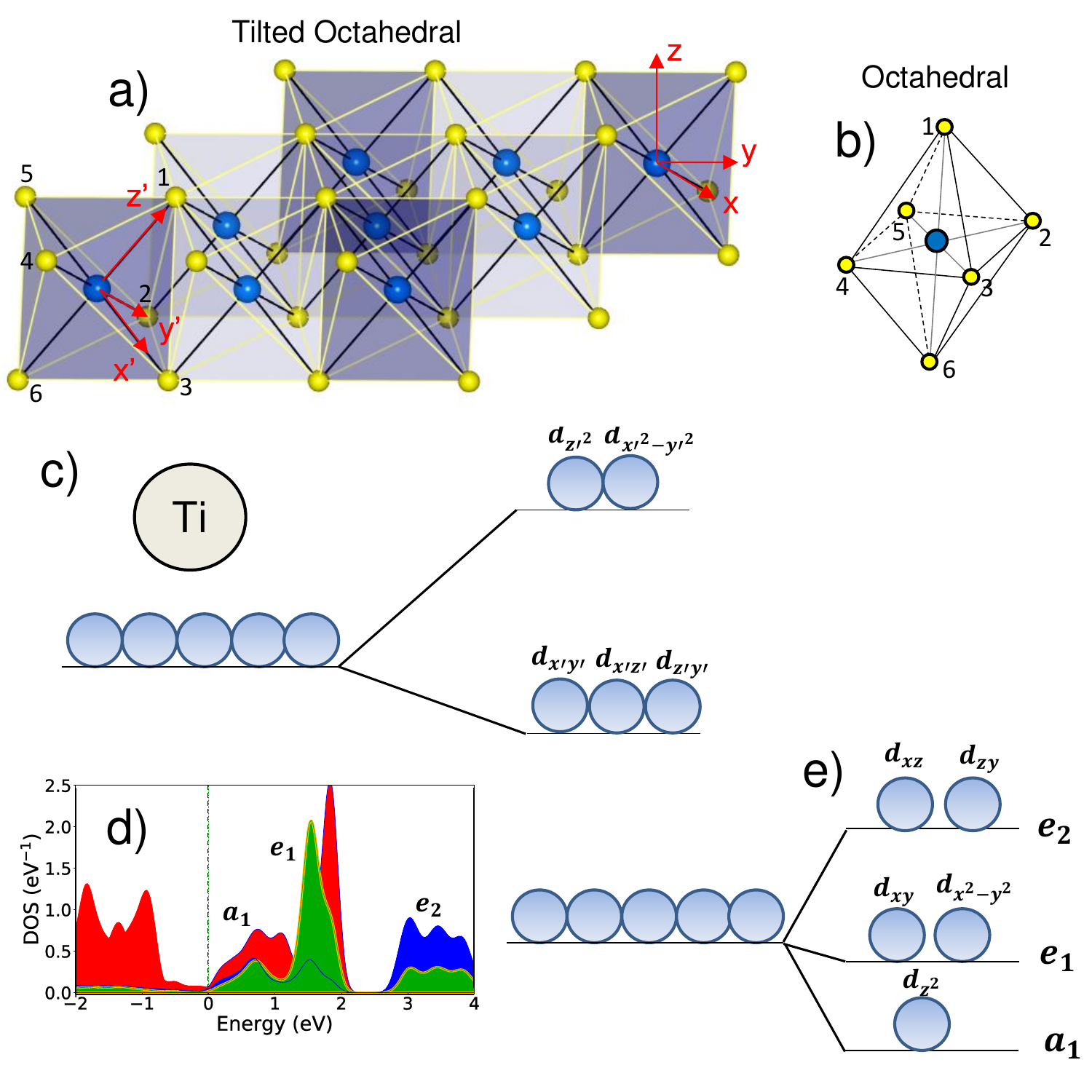}
	\vspace{-0.2 cm}
	\caption{(Color online) (a)  Octahedrons of the single 1T-TiS$_2$ layer. The tilted octahedral axis is in $z'$ direction and one of its faces lie on $xy$ plane.  (b) Spatial arrangement of ligands S in a conventional octahedral structure.
(c) Crystal field splitting of the $3d$ orbitals of Ti atoms in conventional octahedral structure.
The $d$-orbital splits into two different levels. The bottom three energy levels are named  $d_{xy}$,  $d_{xz}$, and  $d_{yz}$ (referred to as  $t_{2g}$ states). The two upper energy levels are named  $d_{x^2−y^2}$, and  $d_{z^2}$ (referred to as  $e_g$ states).                                        
                                          (d) The non-magnetic projected DOS for $d$
orbital of Ti atom in the 1T structure. The Fermi level is set to zero energy.
(e) Crystal field splitting of the $3d$ orbitals of Ti atoms in tilted octahedral structure of 1T-TiS$_2$. 
.}
	\label{fig:subm2}
\end{figure}

\section{Results and discussion}\label{sec3}

\subsection{pristine TiS$_2$}

First, we analyze and compare the electronic properties of a single layer of TiS$_2$ in both 1H and 1T structures. Fig.~\ref{fig:subm1}(a) presents the projected band structure, as well as the density of $3d$-Ti and $3p$-S states for 1H structure. The band structure of 1H-TiS$_2$ reveals a semiconductor nature with an estimated gap of 0.76 eV. Fig.~\ref{fig:subm1}(b) displays the corresponding projected electronic structure for the 1T-TiS$_2$ configuration, where the energy bands intersect the Fermi energy level partially at the $\Gamma$ point. Our spin-polarized calculations show no variation in the density of high and low spin states for these structures, indicating the absence of magnetic properties. Analysis of the band structure and density of states graphs highlights the significant contribution of the $3p$ orbital of S atoms to the valence band region near the $E_F$ in both structures, while the $3d$ orbital of Ti dominates in the conduction band region. A simple model provides insight into the electron filling mechanism, serving as a foundational framework for reviewing all samples in this study. Initially, we examine the electronic configuration of isolated Ti and S atoms. The outer electronic shell of a Ti atom is represented by $4s^2$ $3d^2$, while an S atom is denoted by $3s^2$ $3p^4$. Each sulfur atom requires two electrons to complete its final shell, following the rule of eight. These electrons can be acquired from three neighboring titanium atoms, with each sulfur atom gaining approximately two-thirds electrons from each titanium atom due to its higher electronegativity. Conversely, there are six S atoms surrounding each Ti atom, necessitating each Ti to lose three or four electrons to the six S atoms. According to this model, after bond formation, the $4s$ and $3d$ orbitals of Ti atoms are expected to be nearly empty due to the completion of the $4s^2$ $3d^2$ outer shell. The DOS depicted in Figs.~\ref{fig:subm1}(a) and ~\ref{fig:subm1}(b) corroborate these findings, illustrating a covalent bond between S and Ti atoms through hybridization of the $3p$ orbital of S and the $3d$ orbital of Ti. 

A comprehensive understanding of TMD structures necessitates the consideration of the crystal field effect on the $d$-orbitals of TMs. The significance of the crystal field in the splitting of the $3d$ orbitals within these structures is crucial, particularly when impurities are introduced into the system. When $d$ orbitals are situated in a crystal environment, the degeneracy of the $d$-orbitals is disrupted. In a 1H crystal environment, the five $d$-orbitals are categorized into three groups: two degenerate $d_{xz}$/$d_{yz}$ orbitals, two orbitals $d_{xy}$/$d_{x^2-y^2}$, and single $d_{z^2}$ orbital. 

In 1T structure, the octahedron is not like the conventional perovskite
octahedral structure. The octahedron is tilted with
respect to the standard Cartesian coordinate $xyz$, in
such a way that the 4-fold rotation axes is in the $z'$ direction. 
As shown in Fig.~\ref{fig:subm2}(a), in fact, one of the eight triangles
of octahedron is lying on the floor and the $z$ axis is perpendicular to this triangle.

A conventional octahedral structure is depicted in Fig.~\ref{fig:subm2}(b) and its energy levels are presented in Fig.~\ref{fig:subm2}(c). 
When examining the untilted octahedral crystal 1T structure, it is crucial to consider that the $z$-axis of the conventional octahedron is perpendicular to the monolayer. So, in a rotated primed coordinates $x'y'z'$ , the octahedral crystal environment is conventional, and $d$ orbitals segregate into two groups: $e_g$ ($d_{z'^2}$, $d_{x'^2-y'^2}$) and $t_{2g}$ ($d_{x'y'}$, $d_{x'z'}$, $d_{y'z'}$).
 In the real 1T structure of TiS$_2$, chalcogen atoms [marked 2 to 5 in Fig.~\ref{fig:subm2}(a)] form a plane where the $z$-axis is not perpendicular to this plane.
It is important to note that calculations in this structure were carried out using the $xyz$ coordinate system; thus, the results may not align with those of the primed coordinate system. In the $xyz$ coordinate system, the dual separation of energy states, namely $e_g$ and $t_{2g}$ is not well defined. In this context, our findings are akin to those in reference \cite{Fengyu,Rassekh,Xuedong}, where states are divided into $a_1$, $e_1$, and $e_2$, as illustrated in Fig.~\ref{fig:subm2}(d). This distribution resembles the trigonal prismatic state rather than the octahedral structure as shown in Fig.~\ref{fig:subm2}(e). It means, these outcomes stem from the misalignment of the cartesian coordinate system when dealing with the tilted octahedral structure. Further examination will be conducted when analyzing impurities to provide a more comprehensive analysis.


\subsection{Electronic and magnetic properties in the presence of $3d$ transition metals}

\begin{figure}[b]
	\centering
	\includegraphics[width=75mm]{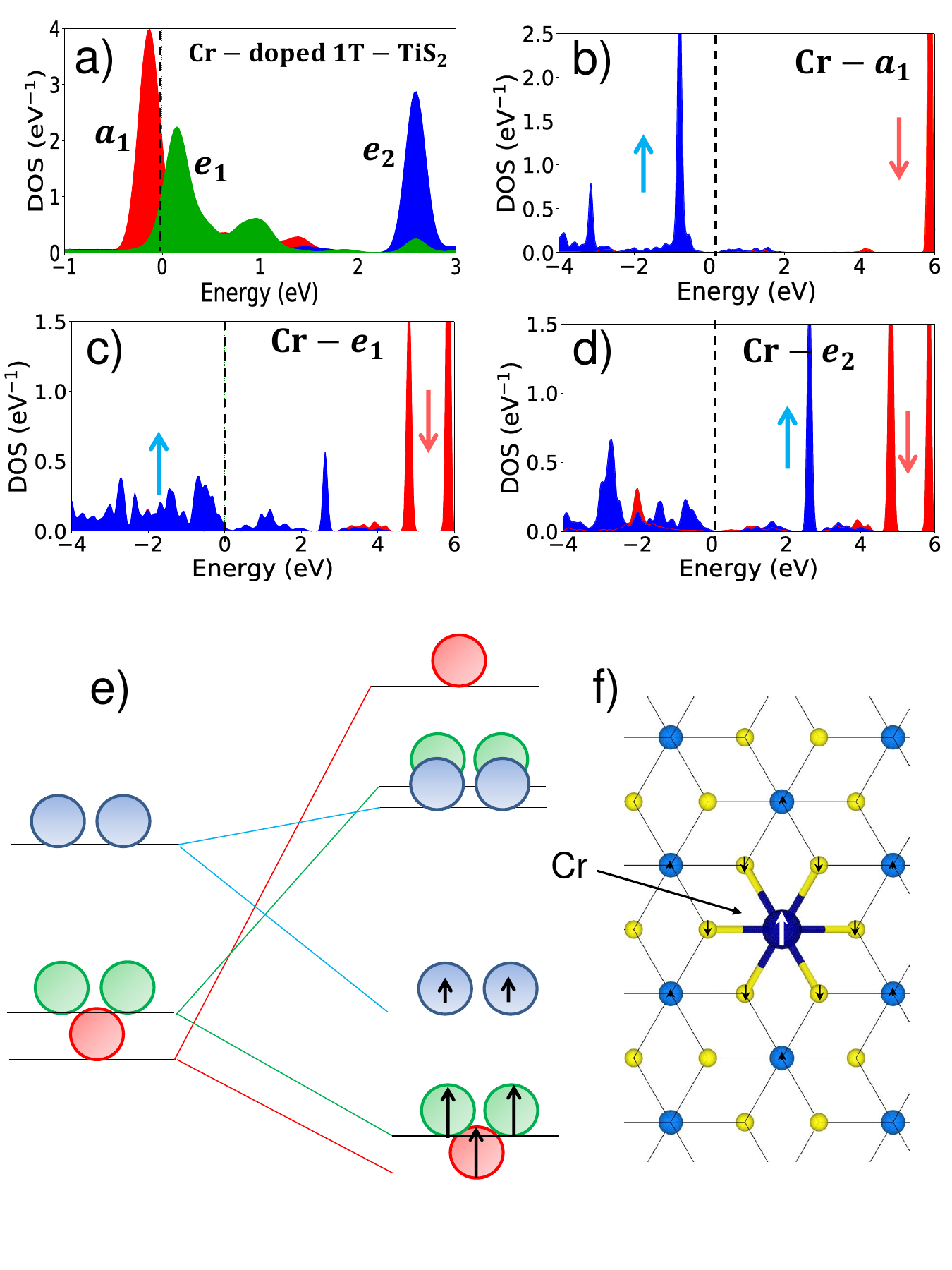}
	\vspace{-0.2 cm}
	\caption{(Color online) (a) The non-spin polarized projected DOS for $d$
orbital of Cr atom in the 1T structure of Cr-doped TiS$_2$. The spin polarized projected DOS in the 1T structure of Cr-doped TiS$_2$ for 
 (b) $d_{z^2}$ ($a_1$), (c) $d_{xy}$/$d_{x^2-y^2}$ ($e_1$), and (d) $d_{xz}$/$d_{yz}$ ($e_2$) orbital characters. The Fermi level is set to zero energy.
(e) Crystal field splitting of the $3d$ orbitals of Cr atoms in Cr-doped tilted octahedral structure of 1T-TiS$_2$ in presence of exchange splitting. (f) Spin polarization around impurity Cr in 1T structure of TiS$_2$. Size of the arrows in the figure indicate the values of spin density distribution in presence of Cr.}
	\label{fig:subm3}
\end{figure}

Now we investigate the impact of substituting $3d$ TMs on the electronic and magnetic properties of both non-magnetic 1H-TiS$_2$ and 1T-TiS$_2$ monolayer structures. Initially, the total energy difference between the magnetic ($E_{sp}$) and non-magnetic ground state ($E_{nsp}$) of the monolayers substitution with $3d$ TMs is calculated. This energy variance is determined using $\Delta E_{spin} = E_{sp}-E_{nsp}$. Table~\ref{table1} presents the total energy difference between the magnetic and non-magnetic ground states of the monolayers following substitution with various TMs. A negative difference signifies a magnetic ground state. The numerical outcomes reveal that $\Delta E_{spin}$ for V, Cr, Mn, Fe, and Cu-doped monolayers is negative in both structures, indicating a magnetic ground state for systems doped with these impurities in both crystal phases. $\Delta E_{spin}$ for Sc and Ni-doped monolayers is zero, suggesting that substituting Sc and Ni does not induce magnetism in either structure. However, the ground state of Co and Zn-doped monolayers is non-magnetic in 1T structure and magnetic in 1H structure. Following the substitution of TM atoms in both structures, a variety of electronic and magnetic phases emerge, including magnetic semiconductor, magnetic semi-metal, non-magnetic semiconductor, non-magnetic metal, and magnetic metal. This structural diversity underscores the significance of this research. Further investigation delves into the magnetic moments of the impurity atom and its first six neighboring atoms. To explore the causes of magnetization in monolayers substituted with TMs, we will initially conduct a comprehensive examination of alloying with Cr atom due to its highest magnetic moment. Subsequently, we will closely analyze the presence of Co atom impurity, which is non-magnetic in the 1T structure and magnetic in the 1H structure, to elucidate the reasons behind this magnetic behavior. Finally, we discuss the rest of the magnetic systems.

\subsection{Cr-doped TiS$_2$}

The numerical values of the magnetic moment of Cr in Table~\ref{table1} suggest that the magnetic properties arise from the introduction of Cr impurities. The presence of Cr atoms as impurities results in the largest magnetic moment among $3d$ TM impurities in both structural phases. From a magnetic perspective, the 1T structure exhibits a higher magnetic moment compared to the 1H structure. To further elucidate the electron configuration in the presence of Cr atoms, we apply a simplified model similar to the one used for the pristine form. Prior to bond formation, the valence shell of the Cr atom consists of $4s^1$ $3d^5$ electrons. Suppose Cr donates the same number of electrons as the Ti atom when it replaces the Ti atom in the structure. With this assumption, and taking into account the full saturation of the sulfur atoms, the expected magnetic moment is approximately 2 $\mu_B$ which is 1-2 $\mu_B$ smaller than the calculated magnetic moments of 3.138 $\mu_B$ for the 1H state and 3.419 $\mu_B$ for the 1T state.

\begin{figure}[t]
	\centering
	\includegraphics[width=86mm]{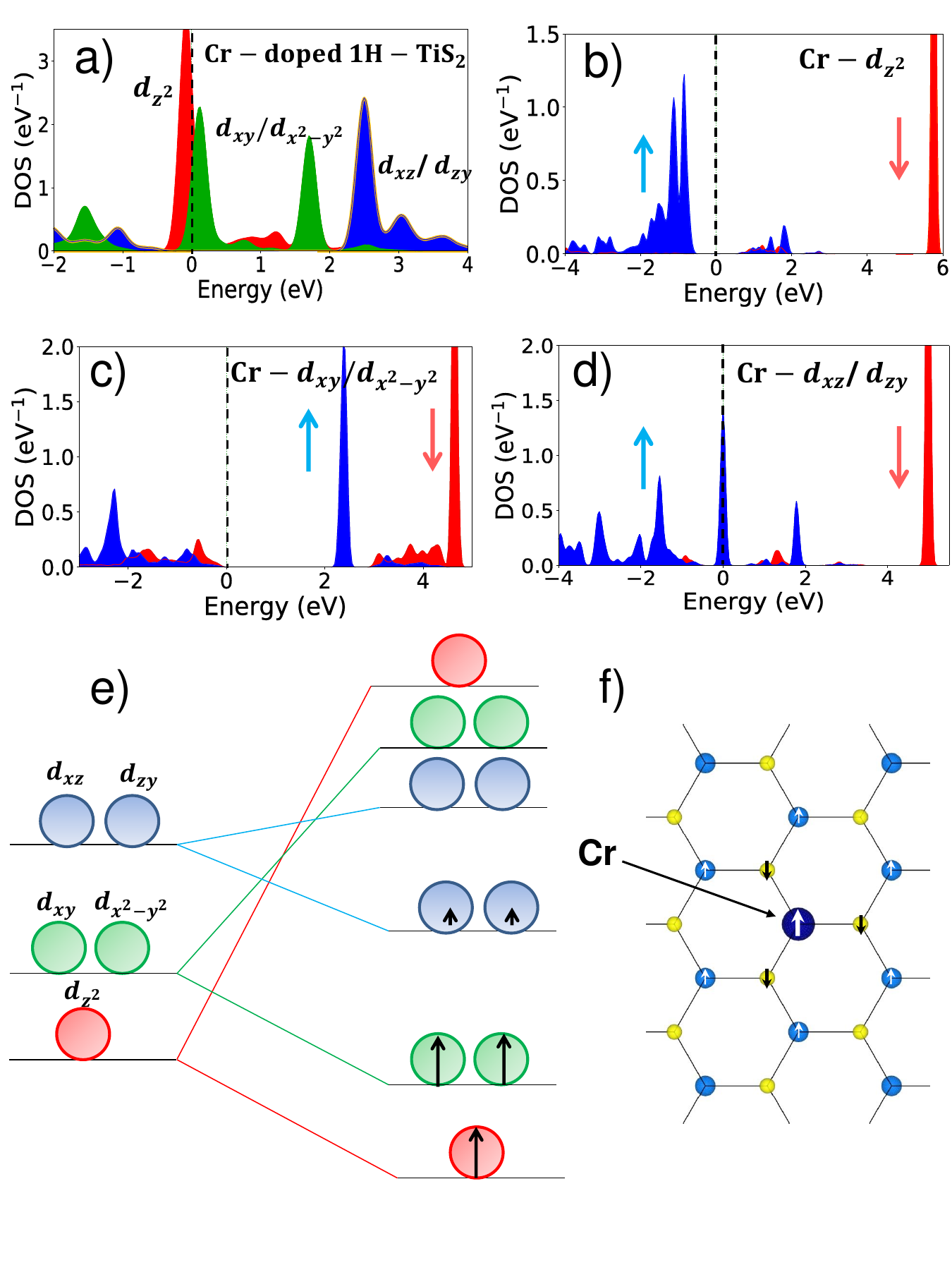}
	\vspace{-0.6 cm}
	\caption{(Color online) (a) The non-spin polarized projected DOS for $d$
orbital of Cr atom in the 1H structure of Cr-doped TiS$_2$. The spin polarized projected DOS in the 1H structure of Cr-doped TiS$_2$ for 
 (b) $d_{z^2}$, (c) $d_{xy}$/$d_{x^2-y^2}$, and (d) $d_{xz}$/$d_{yz}$ orbital characters. The Fermi level is set to zero energy.
(e) Crystal field splitting of the $3d$ orbitals of Cr atoms in Cr-doped tilted octahedral structure of 1H-TiS$_2$ in presence of exchange splitting. (f) Spin polarization around impurity Cr in 1H structure of TiS$_2$. Size of the arrows in the figure indicate the values of spin density distribution in presence of Cr.}
	\label{fig:subm4}
\end{figure}

To provide a clear and accurate explanation of this magnetic moment difference, we delve into two fundamental physical phenomena: the splitting of the $d$ orbitals of the Cr atom due to a crystal field and the spin exchange interaction. The spin exchange interaction involves the alignment of spins of electrons in the $d$ orbitals, contributing to the overall magnetic moment of the Cr atom. We start by using a non-spin calculation with the generalized gradient approximation (GGA) method to examine the splitting of the $3d$ orbitals of the Cr atom in the crystal field created by sulfur ligands in the 1T and 1H structures. 
Contrary to our initial assumption of four electrons loss and complete saturation of neighboring sulfur atoms, Figs.~\ref{fig:subm3}(a) and ~\ref{fig:subm4}(a) show that at least three electrons remain in the $3d$ orbital for chromium atoms in the both 1H and 1T structures. Comparing the occupied states in the 1H and 1T structures reveals that the Cr atom has lost fewer electrons in the 1T state, a trend also observed in the study of Co impurity. This observation could explain the slight increase in magnetic moment in the 1T structure compared to the 1H structure, as discussed further in our analysis.

Our non-spin electronic structures show that the correlated states, including $d_{z^2}$ ($a_1$), $d_{xy}$/$d_{x^2-y^2}$ ($e_1$) in 1H (1T), are partially filled with nearly 3-4 electrons, while the $d_{xz}$/$d_{yz}$ ($e_2$) states are unoccupied. The energy splitting between filled and empty bands, or crystal field splitting, is approximately $E_{CF}$ = 2.2 eV. Cr-doped TiS$_2$ exhibits a large $D(E_F)$ due to its partially filled electronic states with a very small bandwidth. The total bandwidth of correlated subspace in the case of Cr is around $W_b$ = 0.65 eV, significantly smaller than the bandwidth of corresponding states in Ti. The flat bands formed in the presence of these impurities can create the necessary conditions for  correlated phase such as magnetism. Electrons exhibit correlated behavior in this state, and considering the Hubbard $U$ interaction can enhance the accuracy of the calculation. However, the calculation of such complex magnetic ground states using the GGA+$U$ method is beyond the scope of this paper.

To deepen our comprehension of state fillings, calculations incorporating the spin effect should be conducted using the spin-polarized generalized gradient approximation (SGGA) method. The spin exchange interaction plays a critical role in further splitting the levels, laying the groundwork for estimating the high magnetic moment of the Cr atom and similar systems. The outcomes of these calculations are illustrated for the 1T and 1H structures in Figs.~\ref{fig:subm3}(b)-\ref{fig:subm3}(d) and Figs.~\ref{fig:subm4}(b)-\ref{fig:subm4}(d), respectively. Notably, a significant energy gap between the spin-up and spin-down states is evident in both structures when accounting for the spin effects.

We can anticipate a significant exchange splitting in the magnetic calculations due to the sharp peaks of the DOS at $E_F$ in the non-magnetic calculation for the Cr-doped TiS$_2$ compound. As illustrated in Figs.~\ref{fig:subm3} and ~\ref{fig:subm4} for the spin-polarized calculation, the spin splitting of the partially occupied states ranges between  5-6 eV. For example, the energy difference between the occupied spin-up peak and the spin-down peak $a_1$ in 1T structure is 6 eV. In this scenario, the exchange energy $E_{ex}$ and the crystal field splitting $E_CF$ are two competing effects. The Cr-doped TiS$_2$ exhibits a higher exchange energy compared to its crystal field splitting energy, resulting in states with a high spin configuration. The spin-polarized DOS indicates that the number of electrons in the $d$ level is three to four, consistent with the magnetization of 3.1  $\mu_B$ and  3.5 $\mu_B$ as reported in Table~\ref{table1} for 1H and 1T respectively. Additionally, Table I reveals a large magnetic moment of -1.33 $\mu_B$ on the S atoms in close proximity to the impurity in 1H structure, suggesting that the nearest neighbor atoms of the Cr atom are not fully saturated. The magnetization on Ti atoms $m_{Ti}$=0.35 $\mu_B$ suggests that this spin-polarization does not only originate from the Cr in the 1H structure, indicating significant hybridization between impurity Cr and host Zr/S electrons.

\begin{table*}[!ht]
	\caption{Magnetization of impurity  $m_d$, the average magnetization of six S atoms as nearest neighbors of impurity $m_S$,
the average magnetization of the closest Ti to impurity $m_{Ti}$, and spin and non-spin total energy
difference $\Delta E=E_{sp} − E_{nsp}$ for different compounds in 1H and 1T structures.}
	\centering
	\begin{ruledtabular}
		\begin{tabular}{ccccccccccccccc}
			\rule{0pt}{4mm}%
& \multicolumn{2}{c}{$m_d$} &&\multicolumn{2}{c}{$m_S$} && \multicolumn{2}{c}{$m_{Ti}$} &  &\multicolumn{2}{c}{$\Delta E$}
\rule{0pt}{4mm}%
\\ \cline{2-3} \cline{5-6} \cline{8-9} \cline{11-12}
           &1H&1T&&1H&1T&&1H&1T&&1H&1T
\rule{0pt}{4mm}%
\\ \hline
\rule{0pt}{4mm}%
Sc&0.00&0.00&&0.00&0.00&&0.00&0.00&&0.00&0.00\\
V &0.666&0.885&&-0.186&-0.181&&0.489&0.0468&&-76.27&-43.71\\
Cr&3.138&3.419&&-1.332&-0.586&&0.354&0.099&&-1735.54&-3044.98\\
Mn&2.694&2.999&&-0.324&-0.150&&0.663&-0.030&&-603.98&-1175.21\\
Fe&1.838&1.798&&-0.048&-0.018&&0.234&-0.354&&-393.29&-301.60\\
Co&0.805&0.00&&0.09&0.00&&0.084&0.00&&-74.81&0.00\\
Ni&0.00&0.00&&0.00&0.00&&0.00&0.00&&0.00&0.00\\
Cu&0.117&0.345&&0.868&1.002&&-0.154&0.156&&-26.37&-34.02\\
Zn&-0.023&0.00&&1.228&0.00&&-0.24&0.00&&-88.09&0.00\\
		\end{tabular}
		\label{table1}
	\end{ruledtabular}
\end{table*}

\subsection{Co-doped TiS$_2$}

\begin{figure}[t]
	\centering
	\includegraphics[width=87mm]{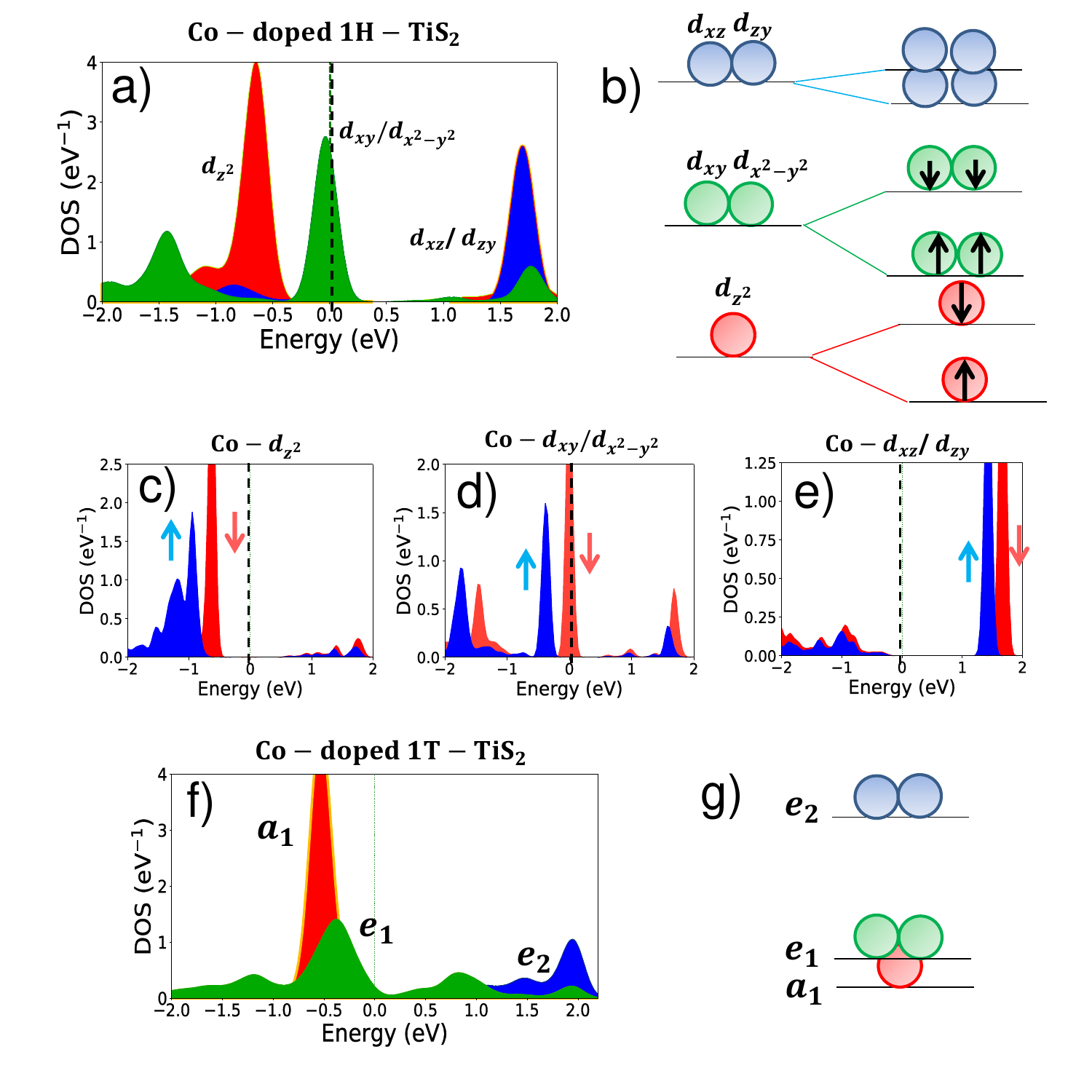}
	\vspace{-0.5 cm}
	\caption{(Color online) (a) The non-spin polarized projected DOS for $d$
orbital of Co atom in the 1H structure of Co-doped TiS$_2$. (b) Crystal field splitting of the $3d$ orbitals of Co atoms in Co-doped tilted octahedral structure of 1H-TiS$_2$ in presence of exchange splitting. The spin polarized projected DOS in the 1H structure of Co-doped TiS$_2$ for 
 (c) $d_{z^2}$, (d) $d_{xy}$/$d_{x^2-y^2}$, and (e) $d_{xz}$/$d_{yz}$ orbital characters. The Fermi level is set to zero energy.
(f) The non-spin polarized projected DOS for $d$
orbital of Co atom in the 1T structure of Co-doped TiS$_2$. (g) Crystal field splitting of the $3d$ orbitals of Co atoms in Co-doped tilted octahedral structure of 1T-TiS$_2$}
	\label{fig:subm5}
\end{figure}

In this section, we will delve into the electronic and spin properties resulting from Co substitution. It's important to note that the motivation behind this study stems from the dual behavior exhibited by Co impurities. With the introduction of this impurity atom, the 1T structure preserves its non-magnetic state, while the 1H structure displays a moderate magnetic moment of around 0.805 $\mu_B$. Initially, the Co atom possesses a $4s^2$ $3d^7$ configuration before bonding. 
To accurately ascertain the electron loss from the $3d$ shell of the Co atom and the orbital arrangement in the presence of the crystal field, the PDOS in both the 1H and 1T structures are depicted in Fig.~\ref{fig:subm5}. It becomes evident from non-spin polarized calculations that there is no distinct separation between $a_1$ and $e_1$ levels in the 1T structure. Conversely, in the crystal field of the 1H structure, there is significant splitting between the $d_{z^2}$ and $d_{xy}$/$d_{x^2-y^2}$ degenerate levels. Furthermore, the presence of half-filled $d_{xy}$ and $d_{x^2-y^2}$ orbitals on the Fermi surface in the 1H structure imparts metallic properties to this state. To comprehend the moderate magnetic ordering in the 1H structure and accurately estimate the electron loss, calculations are conducted while considering the spin effect. The results of these calculations for the 1H structure are illustrated in Figs.~\ref{fig:subm5}(c)-\ref{fig:subm5}(e). The two levels depicted in Fig.~\ref{fig:subm5}(e) are completely devoid of electrons in both spin states. Conversely, the $d_{z^2}$ level is fully occupied in both spin states [see Fig.~\ref{fig:subm5}(c)]. The degenerate $d_{xy}$ and $d_{x^2-y^2}$ levels constitute half-filled states. Notably, spin-up state is located at the $E_F$, resulting in a majority of one spin state metal.
Based on these diagrams, we can infer that there are five electrons in different levels of the cobalt atom, leading to an approximate magnetic moment of 1 $\mu_B$, aligning closely with the reported value of 0.805 $\mu_B$. This arrangement and the new state separations induced by spin-induced exchange interaction are roughly depicted in Fig.~\ref{fig:subm5}(b). 

In Fig.~\ref{fig:subm5}(f) the DOS of the $3d$ orbitals of cobalt impurity atoms in 1T structure is graphed. Accounting for the spin effect, there is zero energy separation $E_{spin}$, indicating a non-magnetic state of the system. As shown in Fig.~\ref{fig:subm5}(f), the $e_1$ and $a_1$ states are fully occupied, while the $e_2$ state remains empty with a significant separation. 
Initially estimated to have six electrons, cobalt in this structure exhibits two potential arrangements as depicted in Fig.~\ref{fig:subm5}. When there is a small gap between the energies of the $e_1$ and $a_1$ states, as illustrated in the Fig.~\ref{fig:subm5}(f), arranging six electrons leads to lack of magnetic ordering. However, if there is a large energy gap between these states, as evident in Fig.~\ref{fig:subm5}(a),  low-spin configuration emerges.

\begin{figure*}[t]
\begin{center}
	\includegraphics[width=150mm]{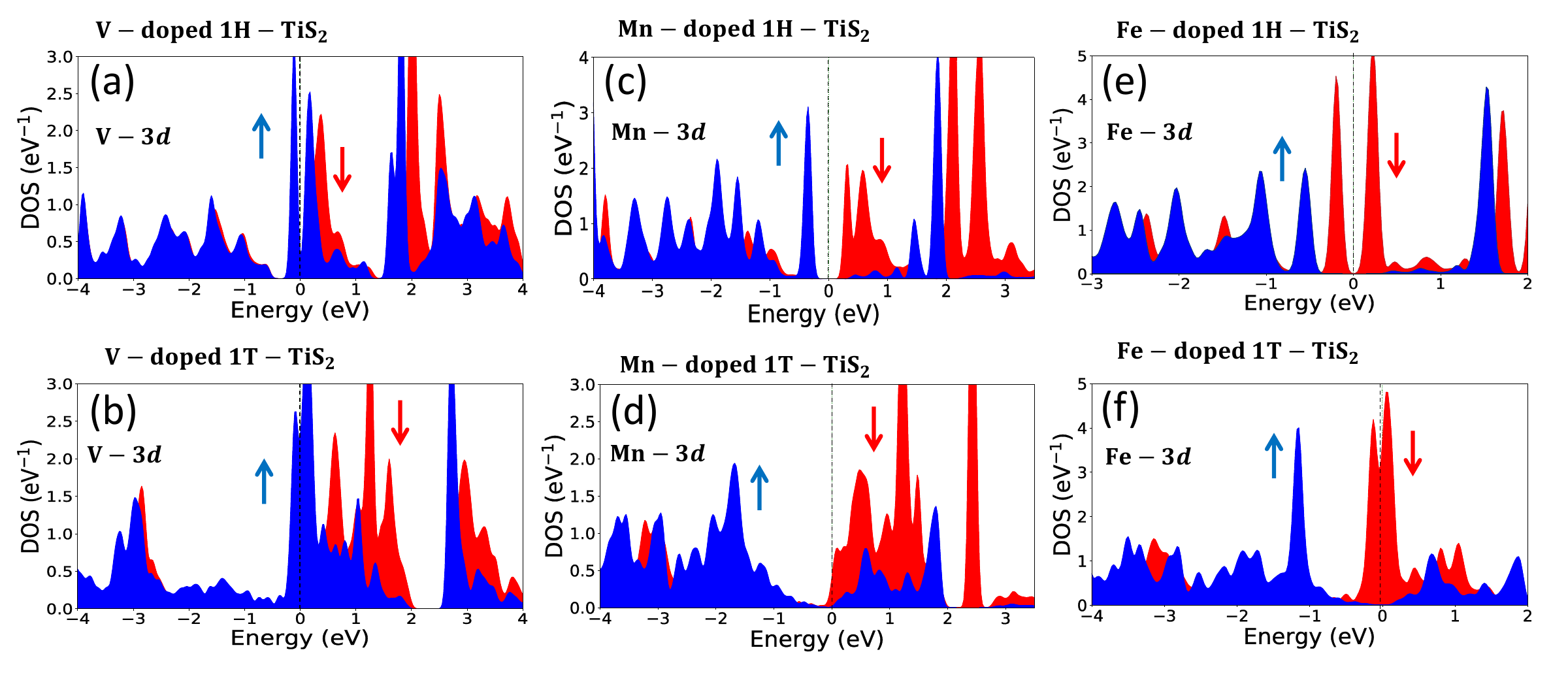}
\end{center}
\vspace*{-0.5cm} \caption{The spin polarized projected DOS of $3d$
orbital of TM impurity atoms in the 1H and 1T structures for  (a), (b) V-doped TiS$_2$; (c), (d) Mn-doped TiS$_2$; (e), (f) Fe-doped TiS$_2$. The Fermi level is set to zero energy.}
\label{fig:subm6}
\end{figure*}

\subsection{V-doped, Mn-doped, and Fe-doped TiS$_2$}

In this section, we explore the substitution of V, Mn, and Fe atoms in TiS$_2$. The V atom carries one fewer electron than Cr, but its magnetic moment is over 2  $\mu_B$ smaller than that of Cr. Within both V-doped structures, the magnetization of neighboring S atoms is small, around -0.18  $\mu_B$. This suggests that the $p$ states of the S atoms are nearly occupied and the V atoms have lost more electrons compared to those in the Cr-doped system. Notably, even the neighboring Zr magnetic moment in the 1T structure drops to zero. The decrease in electron density within the V-doped system diminishes the exchange splitting, as illustrated in Fig.~\ref{fig:subm6}(a) and ~\ref{fig:subm6}(b). Consequently, this leads to a more pronounced crystal field splitting relative to its exchange energy, resulting in states exhibiting a low spin configuration. The presence of electrons with opposite spins serves to weaken the magnetic moment of the impurity and reduce the magnetic moments in the neighboring impurity atoms. In this case, there is no significant difference observed in the magnetization between the 1H and 1T structures, so that they are the magnetic half-metal in both crystal structures.

In general, the magnetization decreases when moving from Cr to heavier atoms. Now, let's discuss the substitution of the Mn atom. 
As illustrated in the 3$d$-Mn DOS plots of Fig.~\ref{fig:subm6}(c) and ~\ref{fig:subm6}(d), the Mn-doped TiS$_2$ compound in the 1H structure manifests as a magnetic semiconductor with a band-gap of approximately 0.30 eV, while it transforms into a magnetic half-metal in the 1T structure. The Mn atom carries one additional electron compared to Cr, and assuming similar charge transfer as in the case of Cr, along with aligned spins, it is expected to exhibit a higher magnetic moment.
Contrary to this expectation, the magnetic moment value of Mn-doped TiS$_2$ is smaller in both structures compared to that of Cr-doped TiS$_2$, measuring at 2.69  $\mu_B$ and 2.99  $\mu_B$ for the 1H and 1T structures, respectively. In contrast to the Cr structure, the magnetization of neighboring S atoms is relatively low, ranging from -0.1  $\mu_B$ to -0.3  $\mu_B$ within the Mn-doped structure.
Within the 1H structure, the exchange splitting value at Mn atoms is smaller than that of the crystal field, resulting in $d$ orbitals adopting a low spin state. In this structure, with three electrons occupying the $d_{z^2}$, $d_{xy}$, and $d_{x^2-y^2}$ levels, the fourth electron with an opposite spin remains at a similar energy level. This spin arrangement maintains the magnetization within the range of 2-3 $\mu_B$.
On the other hand, in the case of the 1T structure, the $d_{xz}$ and $d_{yz}$ orbitals shift to lower energies, diminishing the low-spin state and boosting the magnetic moment of Mn atoms within the 1T structure.

Moreover, by substituting Fe for Ti in TiS$_2$, more than five electrons occupy the Fe-3$d$ levels. These additional electrons result in a magnetization of Fe atoms less than 2 $\mu_B$. Similar to the Mn-doped TiS$_2$ compound, the Fe-doped system in the 1H structure behaves as a magnetic semiconductor, whereas it transitions into a magnetic half-metal in the 1T structure.
.

\section{conclusion and outlook}\label{sec4}

We investigate the electronic and magnetic properties of TiS$_2$ by incorporating V, Cr, Mn, Fe, Co, Ni, Cu, and Zn atoms as impurities in both the 1H and 1T structures using first-principles density functional theory (DFT).
Upon these substituting, a variety of electronic and magnetic phases emerge, including magnetic semiconductor, magnetic half-metal, non-magnetic metal, and magnetic metal.
We focus on two particularly intriguing cases. Introducing Cr atoms as impurities in monolayer TiS$_2$ results in the highest magnetic moment in both the 1T and 1H structures, with the 1T structure exhibiting a slightly larger magnetic moment of 3.419 $\mu_B$ compared to the 1H structure's 3.138 $\mu_B$  attributed to the distorted octahedral structure of the 1T structure. This magnetic half-metal behavior arises from the interplay of two key physical phenomena: the splitting of the TM impurity atom's $d$ orbitals in the crystal field and the spin exchange interaction. In the 1T structure, the Cr atom retains fewer electrons than in the 1H structure due to the distorted octahedral structure. Consequently, Cr substitution can effectively induce magnetism in monolayer TiS$_2$, with the magnetic moment being influenced by the structural phase.
In its pristine form, TiS$_2$ is a non-magnetic semiconductor. The bands near the Fermi energy primarily exhibit $d$ orbital characters, and due to the presence of ideal octahedral and trigonal arrangements, they are well separated from other bands with $p$ character.
Unlike pristine TiS$_2$, the deficiency in saturation of neighboring S atoms in the presence of impurities leads to the proximity of energy levels of $d$ and $p$ states, resulting in unexpectedly sizable magnetic moments even at nearest neighbors to the impurity.

The magnetic behavior of the Co atom demonstrates a significant difference between the two structures. In the 1T structure, the Co atom maintains a magnetic moment of approximately 0.805 $\mu_B$ due to the partial occupation of the degenerate $d_{xy}$ and $d_{x^2-y^2}$ orbitals on the Fermi surface. In contrast, the Co atom in the 1H structure exhibits a non-magnetic state because of the small density of states (DOS) at $E_F$, preventing the formation of a magnetic moment. Our results emphasize the ability to tailor the magnetic properties of Co-doped monolayer TiS$_2$ through structural modifications, with the 1T structure displaying a magnetic state while the 1H structure remains non-magnetic.

Studying the substitution of 3$d$ transition metals in the non-magnetic semiconductor TiS$_2$ offers a fundamental understanding of the origin of magnetic properties in two-dimensional systems.
In the case of Cr/Mn-doped TiS$_2$ in both structures, there exist $d_{z^2}$ and $d_{x^2−y^2}$/$d_{xy}$ impurity levels with a narrow bandwidth of approximately 0.4 eV. The significant DOS of Mn and Cr atoms at the $E_F$ in both the 1T and 1H structures can lead to instability in magnetic ordering, which aligns well with our findings from spin-polarized total energy calculations and the substantial magnetic moments observed.
In non-magnetic calculations, for most of the metallic 3$d$ TM-doped TiS$_2$, we observe substantial $D(E_F)$ values, exceeding those found in elementary transition metals or carbon-based materials. Consequently, the electron correlation strength increases, placing these compounds in the moderately or strongly correlated regime.

\end{document}